\documentclass[compsoc, conference]{IEEEtran}

\usepackage[english]{babel}
\usepackage[rflt]{floatflt}
\usepackage{graphicx}
\usepackage{url}
\usepackage{multirow}
\usepackage[rflt]{floatflt}
\usepackage{graphicx}
\usepackage{multirow}
\usepackage[table]{xcolor}
\usepackage{supertabular}
\usepackage{spverbatim}
\usepackage{paralist}
\usepackage{multicol}
\usepackage{csquotes}
\usepackage{flushend}
\usepackage{cite}

\usepackage{amsmath}
\usepackage{float}
\usepackage{booktabs}
\usepackage[listings]{tcolorbox}
\newcommand\numberthis{\addtocounter{equation}{1}\tag{\theequation}}

\usepackage{enumitem}

\usepackage[inline]{trackchanges}

\addeditor{DS}
\addeditor{RA}
\addeditor{SC}
\addeditor{CM}
\addeditor{ML}

\title{Serverless Protocols for Inventory and Tracking with a UAV}

\begin{document}
\author{\IEEEauthorblockN{Collins Mtita\IEEEauthorrefmark{1}, Maryline Laurent\IEEEauthorrefmark{2}, Damien Sauveron\IEEEauthorrefmark{3},\\
Raja Naeem Akram\IEEEauthorrefmark{4}, Konstantinos Markantonakis\IEEEauthorrefmark{4} and Serge Chaumette\IEEEauthorrefmark{5}
}
\IEEEauthorblockA{\IEEEauthorrefmark{1}TRAXENS S.A.S, Marseille, France\\
\IEEEauthorrefmark{2}SAMOVAR, T\'el\'ecom SudParis, CNRS, Universit\'e Paris-Saclay, \'Evry, France\\
\IEEEauthorrefmark{3}XLIM (UMR CNRS 7252 / Universit\'e de Limoges), D\'epartement Math\'ematiques Informatique. Limoges, France\\
\IEEEauthorrefmark{4}Information Security Group Smart Card Centre, Royal Holloway, University of London, Egham, United Kingdom\\
\IEEEauthorrefmark{5}LaBRI (UMR CNRS 5800 / Universit\'e de Bordeaux), Talence, France \\
Email: c.mtita@traxens.com, maryline.laurent@telecom-sudparis.eu, damien.sauveron@unilim.fr,\\
\{r.n.akram, k.markantonakis\}@rhul.ac.uk, serge.chaumette@labri.fr}}

\maketitle
 
\begin{abstract}
It is widely acknowledged that the proliferation of Unmanned Aerial Vehicles (UAVs) may lead to serious concerns regarding avionics safety, particularly when end-users are not adhering to air safety regulations.  There are, however, domains in which UAVs may help to increase the safety of airplanes and the management of flights and airport resources that often require substantial human resources.  For instance, Paris Charles de Gaulle airport (CDG) has more than 7,000 staff and supports 30,000 direct jobs for more than 60 million passengers per year (as of 2016). Indeed, these new systems can be used beneficially for several purposes, even in sensitive areas like airports.  Among the considered applications are those that suggest using UAVs to enhance safety of on-ground airplanes; for instance, by collecting  (once the aircraft has landed) data recorded  by different systems during the flight (like the sensors of the Aircraft Data Networks - ADN) or by examining the state of airplane structure. In this paper, our proposal is to use UAVs, under the control of the airport authorities, to inventory and track various tagged assets, such as luggage, supplies required for the flights, and maintenance tools. The aim of our proposal is to make airport management systems more efficient for operations requiring inventory and tracking, along with increasing safety (sensitive assets such as refueling tanks, or sensitive pieces of luggage can be tracked), thus raising financial profit.



\end{abstract}

\section{Introduction}

Airlines carry millions of passengers all around the world. This task not only requires to take care of the passengers themselves but also of all necessary equipment, like luggage, tools, refueling vehicles and catering resources.  Tracking all of these is a major challenge that cannot be achieved efficiently by hand, even though this is how it is performed today. From a passenger's perspective, luggage loss is a common occurrence; from an airline safety staff's point of view, a broken or missing safety piece of equipment, e.g. an oxygen mask, can be a major safety risk.

To improve this process, we propose a novel approach where all these assets are the subject of an inventory collated by a UAV. By doing so, we believe that the inventory can be carried out in a timely manner and performed several times if required, depending on flight and spatial constraints of a particular airport.  Naturally, using UAVs in the vicinity of airports is not straightforward.  Authorization by national flight control authorities and airport regulation bodies will be required, but by being situated in an area already subject to strong aerial regulations can make it easier to deal with. Furthermore the different regulation authorities are usually willing today to experiment new real world use cases of UAVs. However, this issue is out of the scope of the current paper.

\subsection{Context}

There exists several initiatives that support the process of luggage management in airports: a) to enable the travelers to track their luggage, e.g. at suitcase manufacturer level~\cite{SamsoniteBeacons2016-04,SamsoniteNBIoT2017-02} or at airline company level~\cite{DeltaRFID2016-05,DeltaRFID2016-11}, from checking/drop-off desk  (and even before, to prevent theft) up to the loading in hold of the aircraft; b) to enable airline companies to improve tracking of checked luggages~\cite{EmiratesRFID2008-02,OptimizationRFID2011-02}.  Moreover, some major airports, such as Hong Kong, Dubai, Las Vegas, have already adopted RFID-enabled baggage handling systems to improve sorting and tracking efficiency.

Our scenario is therefore realistic because it corresponds to a real need and should contribute to improve the process, not only for luggage but also for all the assets airline companies need to deal with (for safety, repair, catering, etc.).  The combination of UAVs and RFIDs for such operations can be more efficient than the standard procedures: UAVs can cover a large area, while embedded RFIDs on aircraft equipment, like oxygen generators, life vests and cabin emergency equipment, can drastically reduce inventory times. Delta Airlines, for instance, has installed more than 240,000 RFID tags on emergency pieces of equipment on all of its own and leased aircrafts. As a result, the company can check their expiration dates aboard an aircraft in a few minutes, rather than approximately eight hours without RFID tags~\cite{DELTARFID2016-09}.

It is already the case that in some airports, not only luggage is equipped with RFID tags, but also maintenance tools~\cite{TAP2012-07,Milano2013-06}. It is thus most likely to become a general approach (perhaps even subject to some regulation) because of the safety enforcement it makes it possible to support.

\subsection{Problem statement}

The context of operation of the UAVs and the goals to achieve raise a number of constraints
and issues that must be taken into account.

\subsubsection{Performance}

From a functional perspective, using UAVs for inventorying implies to two main
requirements. First, the inventory should be achieved in a timely manner, which means that
detecting an RFID and collecting data from it should be as fast as possible. Second, the collection
process and identification protocol should also be efficient to avoid several identification
attempts that would waste time.  Additionally, one should keep in mind that the UAVs
-- at least the kind of small UAV that could be used for this sort of operation -- has limited resources, and the required computation should be minimised.

\subsubsection{Security and privacy}

Security and privacy are strong requirements. In security terms, only the supervising authority should have access to the results of the inventory and tracking processes. Privacy is also important; it should not be possible, for example, for unauthorized persons to trace
pieces of luggage or safety/security-related assets.

Hence, we propose the use of lightweight secure and privacy-preserving serverless protocols as defined in article~\cite{mtita2016efficient} for UAVs and RFIDs (on tagged assets) to perform efficient inventory and search operations. Security is achieved by means of the serverless protocols that enable centrally controlled devices to autonomously authenticate each other without the active participation of a centralized authentication or authorization server~\cite{tan2008secure}. As such, they are appropriate in the airports such that UAVs and RFIDs, even if disconnected from the airport network infrastructure, may establish a mutually authenticated and secure channel among the involved communicating entities ({\em i.e.} among UAVs, and between UAV and RFID tags).

\subsubsection{Energy consumption} In airport contexts, the UAVs ought to be of small size (for safety, security and space reasons), and so energy efficiency is paramount. The rational behind this requirement is based on the limitation of battery size (batteries have a substantial weight and a proper ratio between autonomy and weight has to be found).
Airports being large areas and the number of assets to inventory and track being possibly large, energy consumption must be optimized to ensure a reasonable flight time without reloading the batteries.

\subsection{Contributions}

In this paper, our main goals are to propose two serverless protocols to enhance efficiency in inventory and tracking operations of RFID tagged assets in airport with support of UAVs. 
The proposed protocols, adapted from the Mtita et. al.'s~\cite{mtita2016efficient} protocols for traditional RFID applications, are ideal for the UAVs in the airport environment to ensure reliable and energy-efficient inventory and search operations over some tagged assets without compromising their security and privacy.

The salient contributions of this paper are to propose suitable serverless protocols in the context of airport inventory control and tracking systems with:
\begin{enumerate}[label=\alph*)]
\item an authentication protocol for mass identification of a group of RFID tags within a vicinity;
\item a search protocol to identify a selected group of RFID tags within the proximity.
\end{enumerate}

\subsection{Structure of the Paper}

The remaining of this paper is organized as follows.

Section~\ref{sec:Related Work} presents the related work on UAV-based solutions and serverless protocols.
In Section~\ref{sec:System Model}, we introduce the inventory and tracking system model by describing the involved entities, the requirements to ensure performance (computational and power-consumption efficiency), security and privacy, the assumptions, the threat and attack models.
Section~\ref{sec:Search and Authentication Protocols with UAV scenario} details the two serverless protocols.
In Sections~\ref{sec:Performance Analysis} and~\ref{sec:Security and Privacy Analysis}, the performance, security and privacy analysis of the two protocols are conducted.
Finally, Section~\ref{sec:Conclusions} concludes the paper.

\section{Related Work}
\label{sec:Related Work}
As mentioned previously, there is potential the use of UAVs to facilitate the inventory of assets in airports ({\em e.g.} luggages, supplies required for the flights, maintenance tools) by communicating with the respective tags through authentication and search functionalities. The most feasible way is to make use of RFID technologies: {\em i.e.} attach RFID tags on assets and equip UAV with RFID readers.
The first part of this section presents the inventory UAV-based solution, while the rest is devoted to analyzing RFID-related security protocols.

\subsection{Inventory UAV-based Solutions}
Since their invention, UAVs have been used for surveillance mission: {\em e.g.} for fire detection in forest for civilian application, for enemy detection in military application.
Thanks to their capability to cover wide area in a minimal time, inventory and tracking solutions have been promptly proposed and even some proposals are currently patented. In~\cite{shondel2014unmanned}, Shondel proposed an aerial inventory system for maintaining an inventory record of shipping vessels at a storage facility. In~\cite{mcallister2016mobile}, McAllister claimed invention of a mobile aerial RFID scanning platform.
In~\cite{CarDealership2017-02}, car dealerships claimed to save days of inventory using UAV reading passive ultrahigh-frequency (UHF) RFID tags or BLE beacons attached to cars.
In 2007, Ong et al.~\cite{Ong2007RFID,ong2008mobile} proposed an RFID-equipped UAV to aid inventory automation in a warehouse. Similar ideas were developed by Bae et al. in ~\cite{Bae2016ICISS} and by Andrukiewicz et al. in~\cite{andrukiewicz2016technical}.
In~\cite{Longhi2016emts}, Longhi et al. studied electromagnetic aspects (propagation model, etc.) of the communication between an UAV and passive tags.

Recently, Greco et al.~\cite{Greco2015ICEAA} proposed to use UAV to localize RFID sensors in~\cite{Greco2015rtsi} and to collect data from the RFID sensors scattered throughout the area by simply flying above them.  However, in these two papers, the RFID sensors are not true RFID tags, {\em i.e.} they are not passive tags, but active wireless RFID nodes operating at 433 MHz.
Still related to UAV and RFID, but out of the scope of the paper, in~\cite{Choi2012URAI}, Choi et al. proposed an indoor localization method for UAV using passive UHF RFID tags. It is worth noting that none of the aforementioned work deals with security aspects.

Since there is no work focusing on security and privacy issues between UAV and RFID, the two following papers related to security protocols for UAV and wireless sensors need to be mentioned.
In~\cite{Won2015AsiaCCS}, Won et al. proposed a secure communication protocol enabling a UAV to collect data from smart objects ({\em i.e.} wireless nodes). 
The closest work is the secure and trusted channel protocol proposed by Akram et al.~\cite{Akram-ICNS2017a} to enable in the airport environment a UAV to establish secure communication with sensors of a wireless Aircraft Data Network and other systems to collect data. 
The main difference with these two proposals, apart from the absence of RFID, is that the cryptographic operations used are more complex than those we use in this paper.

\subsection{Serverless Protocols}
\label{sec:related-work}
RFID security protocols can be categorized into two groups: \textit{connection-oriented} and \textit{connectionless} (or \textit{serverless}) protocols. Connection-oriented protocols dictate that an RFID reader -- a UAV in our case -- establishes and maintains a communication channel with the backend or database server during the course of authentication with the tags. Alternatively, connectionless or serverless protocols do not require an established communication between the server and the RFID reader during authentication. The latter case is more pertinent to the UAV case at hand, as it allows for greater mobility and resilience. This section focuses on the serverless authentication protocols, particularly the authentication and secure tag search protocols.

\subsubsection{RFID Authentication Protocols}
\label{auth-prots}
To the best of our knowledge, the use of serverless protocols for RFID authentication was first instigated by Tan et al.~\cite{tan2007severless} in their article~\textit{Serverless search and authentication RFID protocols} published in 2008. They proposed protocols aimed at solving two fundamental problems: first, the identification of tags by readers with no persistent connection to a central database; and second, securely search tags without leaking identifying information to adversaries. In 2013, the authors of~\cite{safkhani2013security} found that Tan et al.'s protocols are vulnerable to traceability, impersonation and privacy attacks. 

In 2009, Lin et al.~\cite{lin2009lightweight} proposed a serverless RFID authentication protocol to improve the computational performance of Tan et al.'s~\cite{tan2007severless} authentication protocol. However, Lee et al.~\cite{lee2012server} note that Lin et al.'s protocol only performs a one sided authentication, that is, the reader authenticates the tag, but the tag does not authenticate the reader. Moreover, Lin et al.'s proposed protocol is still vulnerable to impersonation attack~\cite{lee2012server}.

Hoque et al. proposed a serverless, untraceable authentication, and forward secure protocol for RFID tags~\cite{hoque2010enhancing} claiming that their protocol secures both reader and tag against common attacks with no need for a central database's mediation. But, this claim was disproved by Deng et al.~\cite{deng2014weakness} by showing that Hoque et al.'s protocol was susceptible to data desynchronization attack. Deng et al. also improved Hoque et al.'s~\cite{hoque2010enhancing} authentication protocol in order to withstand data desynchronization attacks. However, the authors of~\cite{pourpouneh2014improvement} found that Deng et al.'s protocol is still vulnerable to data desynchronization attack after two protocol runs. In 2015, Abdolmaleky et al.~\cite{abdolmaleky2015strengthened} proposed a protocol to address the weaknesses found in the protocols proposed by Hoque et al.~\cite{hoque2010enhancing} and Deng et al.~\cite{deng2014weakness}, which are tag impersonation and reader impersonation attacks. Their proposed protocol~\cite{abdolmaleky2015strengthened} solved these problems but after analysis we found that it has very limited use for mass authentication. Indeed, in their proposal, once the reader is granted access to the tag(s), the backend server can no longer access the respective tag(s). This restriction may make sense in some domains of applications, but its usability is very limited in the mass authentication scenarios where disparate readers simultaneously authenticate tags within their vicinity.

ERAP,~\textit{ECC-based RFID Authentication Protocol}~\cite{ahamed2008erap}, is a serverless protocol ensuring mutual authentication between reader and authorized RFID tags. This scheme was found vulnerable to denial of service attack by authors of~\cite{mahalle2012identity}. The authors of~\cite{tang2008efficient} also proposed (HOA)~\textit{HLRO Authentication}, an ECC-based authentication scheme suitable for low-power mobile devices. However this protocol has a strong requirement that each communicating entity has prior knowledge about each other and it is too much CPU and memory demanding as tags must perform ECC and modular operations.

Timestamp is an interesting element for authentication support by constrained devices, as first suggested in 2006 by Tsudik~\cite{tsudik2006trap}. Considering that constrained devices do not have embedded clocks, it was quite a novel idea at the time it was instigated. 
Tsudik's view was simple, a tag stores a static timestamp and an RFID reader periodically broadcasts timestamp of its current time. A tag, in the vicinity of a reader, receives and compares the broadcast timestamp against the stored timestamp. If the broadcast timestamp is larger than the stored timestamp, the tag updates its timestamp and replies with a keyed hash over its permanent key and the new timestamp. Otherwise, the tag sends a random value generated by a Pseudo Random Number Generator (PRNG) to confuse an adversary and avoid narrowing attacks. Narrowing attack occurs when the adversary queries a tag with a particular timestamp and then later tries to identify the same tag by querying a candidate tag with a timestamp slightly above the previous one~\cite{tsudik2006trap}.

Tsudik~\cite{tsudik2006trap} himself noted that his proposed scheme is susceptible to Denial of Service (DoS) attacks as an adversary can easily desynchronize a tag by sending a timestamp value that is ahead of time. This idea was later improved by authors of~\cite{chatmon2006secure} by moving the attack from the resource constrained tag to the powerful backend server.  The improvement aimed at thwarting DoS attacks against the tags but it also resulted to an exhaustive search to the backend server.

The mutual authentication and search protocols adapted in this article were proposed in~\cite{mtita2016efficient}, where the authors claim that their protocols hold in resisting all common security attacks and provide the best performance by using lightweight security primitives.

\subsubsection{Secure RFID Tag Search Protocols}
\label{secure-search}

Like serverless authentication protocols, the idea of secure RFID tag search protocols was introduced by Tan et al.~\cite{tan2007severless} for the purpose of simplifying RFID readers to easily locate a target tag. Nevertheless, tag search protocols have not received the attention~\cite{erguler2016subtle} that mutual authentication protocols received. Nonetheless, they provide a very useful functionality in efficiently locating a specific tag within a group of tags.

Tan et al.~\cite{tan2007severless} proposed tag search protocol. Their protocol is found to perform unidirectional authentication~\cite{chun2011rfid}, i.e the reader authenticates a tag but the tag does not authenticate the reader. In turn, the tag cannot be certain of the authenticity of the reader as any other entity can masquerade as a reader and fool the tag. The protocol is also susceptible to reader's identity disclosure, replay, and impersonation attacks as analyzed by Lee et al.~\cite{lee2012server}. 

In 2009, Lin et al.~\cite{lin2009lightweight} proposed a search protocol by improving Tan et al.'s protocol, but Lin et al.'s protocol was found to be susceptible to replay and impersonation attacks. In 2011, Chun et al.~\cite{chun2011rfid} proposed an RFID tag search protocol with the goal of preserving privacy of communicating parties. However, as the authors of~\cite{he2014secure} noted, Chun et al.'s protocol is susceptible to tracking attacks due to static values sent from the reader to the tag. 

In 2012, Lee et al.~\cite{lee2012server} also proposed an RFID search protocol. Their protocol uses hash function twice on the same parameter and also makes use of PRNG on the tag side. These operations consume a lot of resources with respect to the computational constraints of most RFID tags~\cite{he2014secure}.

In 2014, Xie et al.~\cite{xie2014rfid} proposed a secure tag search protocol in their article, \textit{RFID seeking: Finding a lost tag rather than only detecting its missing}, which is secure against common
attacks such as replay, traceability and DoS. Xie et al.'s~\cite{xie2014rfid} protocol was later improved by Jeon et al.~\cite{jeon2014ultra} in 2014. Jeon et al.'s protocol suffers from the reader traceability attack, which was not in the original protocol proposed by Xie et al.~\cite{xie2014rfid}. It is observed that the lack of context in the protocol between the reader and the tag leads to the replay attacks.

In 2017, Sundaresan et al.~\cite{sundaresan2017secure} proposed a secure search protocol for low cost passive RFID tags, which is based on quadratic residues. The authors claim that the tag running their protocol performs only simple security primitives such as $XOR$ ($\oplus$), modular arithmetic operation ($mod$) and 128-bit PRNG, thus achieving compliance with EPC standards. However, the author specifically state that their protocol requires the reader must maintain a connection to the back-end server during authentication phase as the server must perform some of the critical operations; this disqualifies it as a candidate for search protocols for the scenarios in this article.

The search protocols described above make use of static authentication parameters, which do not expire. This implies that the reader is only authorized once and the parameters remain valid forever. Moreover, once the reader is compromised, the parameters cannot be revoked, hence tags can be accessed by adversaries without any remedies to the problem.

In 2016, Mtita et al.~\cite{mtita2016efficient} proposed a secure serverless search protocol which is adapted in this article to provide a secure search functionality for the UAVs. The authors claim that their protocol is lightweight, as tags perform very few operations during the search query and only one tag responds, if the right tag is present. Likewise, they claim that the search protocol is resistant to narrowing attacks, replay and cloning attacks~\cite{tsudik2006trap}.

\section{Inventory and Tracking System Model}
\label{sys-model}
\label{sec:System Model}
This section outlines specifications for each player involved in the system and protocols,  in addition to the performance, security and privacy requirements, assumptions, and threat and attack models.
\subsection{Entities}
The protocols proposed in this article involve the interaction between three parties as presented below with their respective characteristics. The definition of each parameter used in the protocols is provided in Table~\ref{table:params}.
\begin{compactitem}
\item \textbf{Backend Server}: denoted as $S$ is a trusted, powerful entity with unlimited resources. $S$ has a list of all legitimate tags and UAVs, hence it plays a role of assigning parameters to UAVs for accessing authorized tags. Note that the server is offline when the UAV is launching an authentication session with the tags. 
\item \textbf{Unmanned Aerial Vehicle (UAV)}: denoted as $UAV_j$, has finite resources for storage, computation, energy and communication. $UAV_j$ stores a list of tags $L_j$, which represents all authorized tags that $UAV_j$ can authenticate and exchange information with. $UAV_j$ remains untrusted by the tags until the mutual authentication phase is successfully completed.
\item \textbf{RFID Tag}: denoted as $\rho_i$, the tag is characterized by scarce resources in terms of storage, computation, energy and communication. Each tag $\rho_i$ has a unique identifier $id_i$ that doubles as a secret key shared with the backend server $S$. Likewise, each tag $\rho_i$ has a static timestamp $T_{SYS}$ initialized at the time of tag's manufacture and does not need to be tag-unique.
\end{compactitem}

\subsection{Performance Requirements}
To ensure computational and energy consumption efficiency of our protocols, the main players of the protocols, i.e. UAVs and tags, must only use lightweight operations. 

\subsection{Security and Privacy Requirements}
The following security and privacy requirements must be present in our proposed mutual authentication and secure tag search protocols. 

\subsubsection*{Mutual Authentication}
Our protocols must perform mutual authentication in order to establish mutual trust between tags and UAVs, eventually avoiding impersonation.

\subsubsection*{Freshness}
Protocols must enforce message freshness in order to thwart replay attacks. Our proposed protocols enforce freshness by generating each message using random values during each protocol run.

\subsubsection*{Untraceability}
Non-traceability, or untraceability, entails that it should not be possible for a tag (or UAV) to be identified based solely on the exchanged messages nor to link two different sessions to the same tag (or UAV).

\subsection{Assumptions}
\label{assumptions}
\begin{compactitem} 
\item Backend server is a trusted entity and cannot be compromised. 
\item Backend server allocates only a fraction of RFID tags to each UAV for authentication.
\item PRNG and keyed Hash-based Message Authentication Code (HMAC) functions are considered as robust.
\end{compactitem}

\subsection{Threat and Attack Models}
\label{threat-models}
\begin{figure*}[ht!]
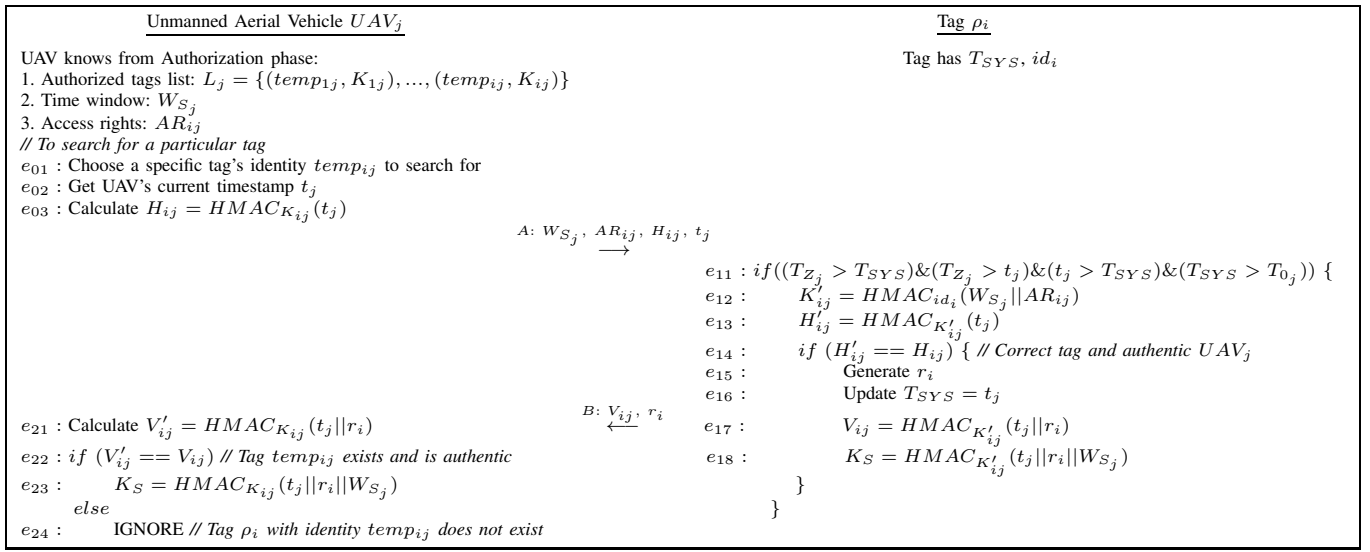

\centering
\begin{scriptsize}
\fbox{
\begin{minipage}{0.99\textwidth}
\hspace{1.5cm}~ \underline{Unmanned Aerial Vehicle $UAV_j$}\hspace{7cm} \underline{Tag $\rho_i$}
\begin{flushleft}
UAV knows from Authorization phase:   \hspace{7.6cm} Tag has $T_{SYS}$, $id_{i}$ 
\\1. Authorized tags list: $L_j = \{(temp_{1j}, K_{1j}), ..., (temp_{ij}, K_{ij})\}$
\\2. Time window: $W_{S_j}$
\\3. Access rights: $AR_{ij}$\\
\textit{// To search for a particular tag}\\
$e_{01}:$ Choose a specific tag's identity $temp_{ij}$ to search for\\
$e_{02}:$ Get UAV's current timestamp $t_j$\\
$e_{03}:$ Calculate $H_{ij} = HMAC_{K_{ij}}(t_j)$\\  

\hspace{6.5cm} $ \overset{A:~W_{S_j},~AR_{ij},~H_{ij},~t_j}{\longrightarrow}$ 
\\                                                                                          
\hspace{9.1cm}$e_{11}:$  $if( (T_{Z_j}>T_{SYS}) \& (T_{Z_j}>t_j) \&(t_j>T_{SYS})\&(T_{SYS}>T_{0_j}) )$ \{\\
\hspace{9.1cm}$e_{12}:$\hspace{0.6cm} ${K'_{ij}} = HMAC_{id_i}(W_{S_j}||AR_{ij})$\\ 
\hspace{9.1cm}$e_{13}:$\hspace{0.6cm} $H'_{ij} = HMAC_{K'_{ij}}(t_j)$\\
\hspace{9.1cm}$e_{14}:$\hspace{0.6cm} $if$ $(H'_{ij} == H_{ij}) $ \{ \textit{// Correct tag and authentic $UAV_j$}\\
\hspace{9.1cm}$e_{15}:$\hspace{1.2cm} Generate $r_{i}$\\
\hspace{9.1cm}$e_{16}:$\hspace{1.2cm} Update $T_{SYS} =  t_j$\\
$e_{21}:$ Calculate $V'_{ij} = HMAC_{K_{ij}}(t_j || r_{i})$ ~\hspace{2.5cm} $\overset{B:~ V_{ij},~r_{i}}{\longleftarrow}$ \hspace{0.35cm} $e_{17}:$\hspace{1.2cm} $V_{ij} = HMAC_{K'_{ij}}(t_j || r_{i})$\\
$e_{22}:$ $if$ $(V'_{ij} == V_{ij}) $ \textit{// Tag $temp_{ij}$ exists and is authentic}~\hspace{2.4cm} $e_{18}:$\hspace{1.2cm} $K_S = HMAC_{K'_{ij}}( t_j || r_{i} || W_{S_j})  $\hspace{0.2cm}~\\
$e_{23}:$\hspace{0.6cm} $K_S = HMAC_{K_{ij}}( t_j || r_{i} || W_{S_j}) $\hspace{5.1cm}~ \}\\
\hspace{0.7cm}$else$ \hspace{8.6cm}~\}\\
$e_{24}:$\hspace{0.6cm} IGNORE \textit{// Tag $\rho_i$ with identity $temp_{ij}$ does not exist}
\end{flushleft}
\end{minipage}
}
\caption{Our Serverless Secure RFID Tag Search Protocol }
\label{search-protocol}
\end{scriptsize}
\end{figure*}

To model the security and privacy for our protocols, we consider a polynomial time adversary $\alpha$ attacking our proposed protocols following the games described below with the aim of gaining access to secret information or disrupting a normal protocol run. The security and privacy games are designed to show the capabilities, limitations and options of the adversary as he attempts to break the protocols. The games described hereafter can apply to the proposed protocols depicted in Figures~\ref{authentication-protocol} and~\ref{search-protocol}. 
\\\textit{\textbf{Game 1}: $\alpha$ masquerades as UAV}
\begin{compactitem}
\item \textbf{step 1.1}: $\alpha$ observes and eavesdrops several exchanges between legitimate $UAV_j$ and one or more tags. 
\item \textbf{step 1.2}: $\alpha$ sends messages~\textit{A} and~\textit{C} (respectively, only message~\textit{A} for the tag search protocol) to tag $\rho_i$.\\$\alpha$ wins if he can send valid message~\textit{C}.
\end{compactitem}
\textit{\textbf{Game 2}: $\alpha$ creates a new counterfeit tag $\rho_x$}
\begin{compactitem}
\item \textbf{step 2.1}: $\alpha$ physically attacks $\rho_i$'s to access its data. 
\item \textbf{step 2.2}: $\alpha$ uses the data from valid $\rho_i$ to create other counterfeit tags $\rho_x$ where $x \neq i$.\\$\alpha$ wins if he can create counterfeit tag $\rho_x$ and fool legitimate $UAV_j$.
\end{compactitem}
\textit{\textbf{Game 3}: $\alpha$ tracks tag $\rho_i$}
\begin{compactitem}
\item \textbf{step 3.1}: $\alpha$ is able to observe exchanges between legitimate $UAV_j$ and tags $\rho_1$ and $\rho_2$, one after the other, for a polynomial number of times each.
\item \textbf{step 3.2}: The challenger selects a tag $\rho_i$, $i \in \{1, 2\}$, and let it authenticate to $UAV_j$. $\alpha$ listens to the exchanges and sends a guessed value $i$ to the challenger. \\
$\alpha$ wins the game if value $i$ is correct. The protocol is considered private if $\alpha$ cannot win the game with a probability greater than 0.5.
\end{compactitem}


\section{Search and Authentication Protocols}
\label{sec:Search and Authentication Protocols with UAV scenario}

This section presents security protocols relevant to securing the communication between UAVs and the corresponding authorized tags.

Due to the high mobility of UAVs, serverless mutual authentication and search protocols seem ideal to solve the security and privacy problems. A mutual authentication protocol helps to simultaneously authenticate an UAV with a large number of tags attached to assets ({\em e.g.} luggages, supplies required for the flights, maintenance tools). The mass authentication is useful where a large number of assets need to be securely and quickly authenticated at once.

On the other hand, the secure search protocol is useful in efficiently locating a specific tag attached to a baggage among a number of other tags. The efficiency of the search protocol is due to its ability to narrow down the query that forces only the target tag to respond to the authentic request.


The two serverless protocols proposed in this section complement one another and share the same first phase. The first phase presented in section~\ref{authentication-initialization} involves authorization between an UAV and the central backend server. Each UAV must perform this phase prior to commencing the second phase of the protocols, which involves either authentication or search. We describe the common phase before we start explaining how each of the individual authentication and search phase work. 

\subsection{Protocol Notations}
\label{notation}

The protocols description and figures in the following sections will be described using  notations given in Table~\ref{table:params}. 

In Table~\ref{table:params}, $AR_{ij}$ represents  encoded access rights for $UAV_j$ with respect to the data stored in tags $temp_{ij}$. In this article, $AR_{ij}$ is represented in the form of a code, like Unix file permissions, with~\textit{Read, Write} and~\textit{Execute} options. Time window $W_{S_j}$ is a 64-bit parameter represented as $[T_{0_j}||T_{Z_j}]$, where $T_{0_j}$ is the start date and $T_{Z_j}$ is the end date defining the time limits for the specific $UAV_j$ to access the tags within the list $L_j$.

\begin{table}[H]
\centering
\begin{scriptsize}
\caption{Protocol notations with size estimations}
\begin{tabular}{llr}
\toprule
Parameter name & Symbol & Bits \\
\midrule
Tag's Static Timestamp & $ T_{SYS} $ &32\\
UAV's Timestamp & $ t_j$ &32\\
Start date	& $T_{0_j}$	&32\\
End date	& $T_{Z_j}$	&32\\
Time window	& $W_{S_j}$	&64\\
Access rights & $AR_{ij}$ &128\\
Tag's Random Number & $r_{i}$ & 128\\
UAV's Random Number & $r_j$ & 128 \\
Tag's Identifier&	$id_i$ & 	128\\
Temporary Tag's Identifier&	$temp_{ij}$ & 	128\\
Tag's Key&	$K_{ij}$ & 	160\\
HMACs (from UAV or Tag) &	$H_{ij}$, $V_{ij}$	&160\\
Identify of $UAV_j$ &	$ID_{UAV_{j}}$	&-\\
List of authorized tags &	$L_{j}$	&-\\
Concatenation operator & $||$ & -\\
\bottomrule
\end{tabular}
\label{table:params}
\end{scriptsize}
\end{table}

\begin{figure*}[ht!]
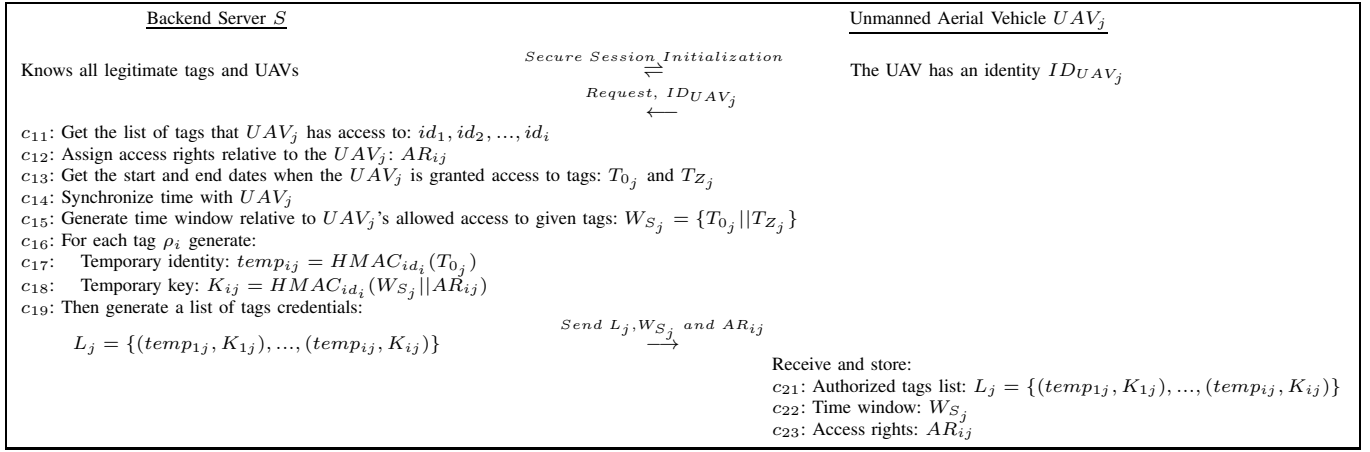

\centering
\begin{scriptsize}
\fbox{
\begin{minipage}{.99\textwidth}
\hspace{1.5cm}~ \underline{Backend Server $S$}\hspace{7.4cm} \underline{Unmanned Aerial Vehicle $UAV_j$}
\begin{flushleft}
Knows all legitimate tags and UAVs \hspace{2.9cm}$ \overset{Secure~Session~Initialization}{\rightleftharpoons} $\hspace{0.8cm} The UAV has an identity $ID_{UAV_{j}}$\\
\hspace{7.5cm}$ \overset{Request,~ID_{UAV_{j}}}{\longleftarrow}$\\
$c_{11}$: Get the list of tags that $UAV_j$ has access to: $id_1, id_2, ..., id_i$ \\
$c_{12}$: Assign access rights relative to the $UAV_j$: $AR_{ij}$\\
$c_{13}$: Get the start and end dates when the $UAV_j$ is granted access to tags: $T_{0_j}$ and $T_{Z_j}$\\
$c_{14}$: Synchronize time with $UAV_j$\\
$c_{15}$: Generate time window relative to $UAV_j$'s allowed access to given tags:  $W_{S_j} = \{T_{0_j}||T_{Z_j}\}$ \\
$c_{16}$: For each tag $\rho_i$ generate:\\
$c_{17}$: ~~~Temporary identity: $temp_{ij} = HMAC_{id_i}(T_{0_j})$\\
$c_{18}$: ~~~Temporary key: $ K_{ij} = HMAC_{id_i}(W_{S_j}||AR_{ij}) $\\
$c_{19}$: Then generate a list of tags credentials:\\
~~~~~~~~$L_j = \{(temp_{1j}, K_{1j}),..., (temp_{ij}, K_{ij})\} \hspace{1.5cm}  \overset{Send~ L_j, W_{S_j}~ and~ AR_{ij}}{\longrightarrow}$ \hspace{0.4cm} \\
\hspace{9.9cm} Receive and store: \\
\hspace{9.9cm} $c_{21}$: Authorized tags list: $L_j = \{(temp_{1j}, K_{1j}),..., (temp_{ij}, K_{ij})\}$\\
\hspace{9.9cm} $c_{22}$: Time window: $W_{S_j}$\\
\hspace{9.9cm} $c_{23}$: Access rights: $AR_{ij}$ 
\end{flushleft} 
\end{minipage}
}
\caption{Authorization between a backend RFID server and a UAV supporting authentication and access rights assignment}
\label{authentication-protocol-session1}
\end{scriptsize}
\end{figure*}

\subsection{Authorization between Backend Server and UAV}
\label{authentication-initialization}

The~\textit{authorization phase}, depicted in Figure~\ref{authentication-protocol-session1}, involves the exchange between an UAV, $UAV_j$, and the backend server, $S$, through a secure channel, where $UAV_j$ acquires appropriate access rights from the server to access a group of tags attached to assets.

\begin{compactenum}
\item $S$ generates a key $K_{ij}$ and a temporary identity $temp_{ij}$ corresponding to each tag $\rho_i$ that $UAV_j$ is authorized to access to with the given access rights $AR_{ij}$. The key $K_{ij}$ and identity $temp_{ij}$ of each tag are ephemeral and derived from the time window $W_{S_j}$ and start date $T_{0_j}$ generated by $S$, respectively. 
\begin{align*}
 K_{ij} = HMAC_{id_i}(W_{S_j}||AR_{ij}) \numberthis \label{eqn01}
\end{align*}
\begin{align*}
 temp_{ij} = HMAC_{id_i}(T_{0_j}) \numberthis \label{eqn03}
\end{align*}
\item $S$ builds a list of authenticated tags $L_j$ granted to $UAV_j$ for a given time window $W_{S_j}$ with access rights $AR_{ij}$. $S$ is assumed to assign different time windows $W_{S_j}$ and $AR_{ij}$ to different UAVs.  
\begin{align*}
 L_j = \{(temp_{1j}, K_{1j}), ..., (temp_{ij}, K_{ij})\}~~ \numberthis \label{eqn02}
\end{align*}
\item $S$ securely sends $L_j$, $AR_{ij}$, and $W_{S_j}$ to $UAV_j$.
\end{compactenum}

\begin{figure*}[ht!]
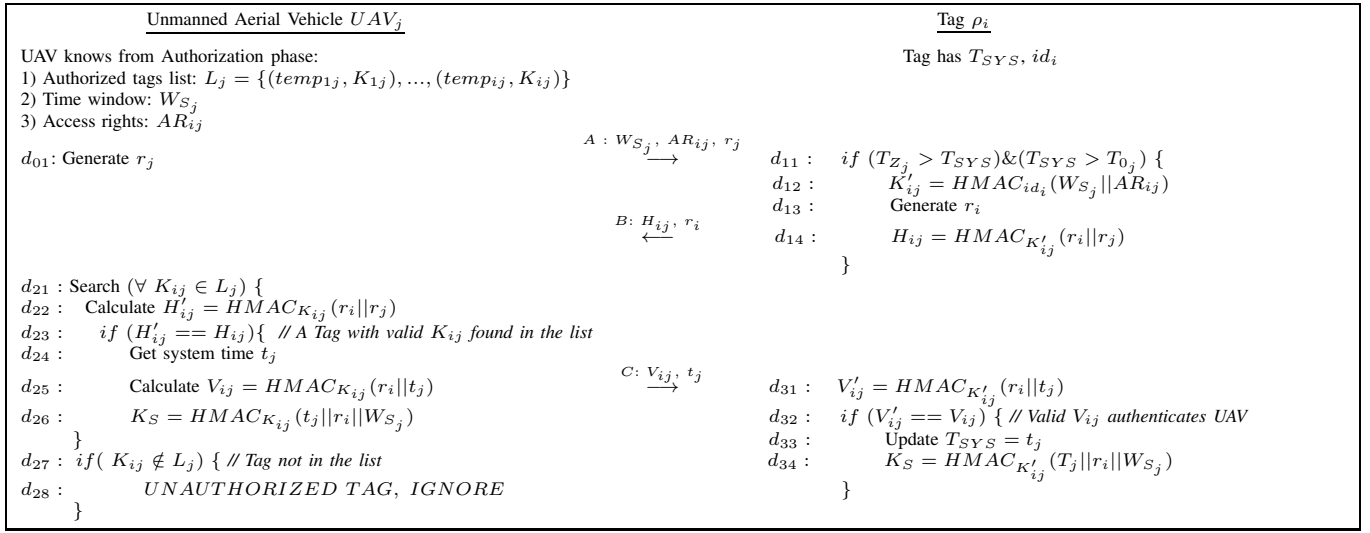

\centering
\begin{scriptsize}
\fbox{
\begin{minipage}{0.99\textwidth}
\hspace{1.5cm}~ \underline{Unmanned Aerial Vehicle $UAV_j$}\hspace{7cm} \underline{Tag $\rho_i$}
\begin{flushleft}
UAV knows from Authorization phase: \hspace{7.60cm} Tag has $T_{SYS}$, $id_i$\\
1) Authorized tags list: $L_j = \{(temp_{1j}, K_{1j}), ..., (temp_{ij}, K_{ij})\}$\\
2) Time window: $W_{S_j}$\\ 
3) Access rights: $AR_{ij}$

$d_{01}$: Generate $r_j$ \hspace{5.5cm} $ \overset{A~:~W_{S_j}, ~AR_{ij}, ~r_j}{\longrightarrow}$ \hspace{.2cm} $d_{11}: \hspace{0.3cm} if~(T_{Z_j}>T_{SYS})\&(T_{SYS}>T_{0_j})$ \{~\hspace{0.3cm}~\\
\hspace{9.9cm} $d_{12}: \hspace{0.9cm}{K'_{ij}} = HMAC_{id_i}(W_{S_j}||AR_{ij})$\\ 
\hspace{9.9cm} $d_{13}:$ \hspace{0.84cm} Generate $r_{i}$ \\
\hspace{7.8cm} $\overset{B:~ H_{ij},~r_{i}}{\longleftarrow}$ \hspace{.8cm} $d_{14}: \hspace{0.9cm} H_{ij} = HMAC_{K'_{ij}}(r_{i}||r_j)$\\
\hspace{10.9cm}\}

$d_{21}:$ Search $( \forall~K_{ij} \in L_j)~\{$\\
$d_{22}:$\hspace{0.3cm}Calculate $H'_{ij} = HMAC_{K_{ij}}(r_{i}||r_j)$\\
$d_{23}:$ \hspace{0.4cm}$if$ $(H'_{ij} == H_{ij}) \{$ ~\textit{//~A Tag with valid $K_{ij}$ found in the list}\\
$d_{24}:$ \hspace{0.8cm}Get system time $t_j$\\
$d_{25}:$ \hspace{0.8cm}Calculate $V_{ij} = HMAC_{K_{ij}}(r_{i} || t_j)$ \hspace{2.3cm} $\overset{C:~ V_{ij}, ~t_j}{\longrightarrow}$ \hspace{0.4cm}  \hspace{0.2cm} $d_{31}:~~~ V'_{ij} = HMAC_{K'_{ij}}(r_{i} || t_j)$\\
$d_{26}:$\hspace{0.8cm}  $K_S =  HMAC_{K_{ij}}( t_j || r_{i} || W_{S_j})  $ \hspace{4.54cm} $d_{32}:$\hspace{0.3cm} $if~(V'_{ij} == V_{ij})~\{$ \textit{//~Valid $V_{ij}$ authenticates UAV}\\ 
\hspace{0.6cm}~$\}$ \hspace{8.94cm} $d_{33}:\hspace{0.9cm}$ Update $T_{SYS} = t_j$ \\
$d_{27}:$  $~if (~K_{ij} \notin L_j)~ \{$~\textit{// Tag not in the list} \hspace{5.04cm}$d_{34}:$ \hspace{0.9cm}$K_S =  HMAC_{K'_{ij}}( T_{j} || r_{i} || W_{S_j})  $\\
$d_{28}:$ \hspace{0.9cm}  $UNAUTHORIZED ~TAG,~IGNORE$  \hspace{4.3cm} $\}$    \\
~~~~~~~~$\}$
\end{flushleft}
\end{minipage}
}\caption{Our Serverless Authentication Protocol between UAV and Tag}
\label{authentication-protocol}
\end{scriptsize}
\end{figure*}

\subsection{Serverless Authentication between UAV and Tags}
After running a mandatory preliminary phase of authorization with the backend server, depicted in section~\ref{authentication-initialization} and Figure \ref{authentication-protocol-session1}, $UAV_j$ is ready to perform mutual authentication phase with the tags in the list $L_j$. The mutual authentication phase, described in Figure~\ref{authentication-protocol}, involves verification and authentication between $UAV_j$ and a~\textit{tag} $\rho_i$ with the purpose of guaranteeing the authenticity of UAVs and tags during communication and exchange of secret data. 


As $UAV_j$ flies over assets, it broadcasts a message $A$ containing $W_{S_j},~AR_{ij}$, and $r_j$. All tags within the vicinity of $UAV_j$ respond with a challenge containing $r_i$ and $H_{ij} = HMAC_{K'_{ij}}(r_i||r_j)$, where $r_i$ is the random number generated by a respective tag. Upon receipt of message $B$ from multiple tags, $UAV_j$ calculates $H'_{ij} = HMAC_{K_{ij}}(r_i||r_j)$ using the values of $K_{ij}$ in the list $L_j$. If the corresponding value of $H_{ij}$ is found, $UAV_j$ authenticates the respective tag and replies with $V_{ij}$ and $t_j$ via message $C$. 

Upon receipt of message $C$, $\rho_i$ checks the validity of $V_{ij}$. The correct value of $V_{ij}$ authenticates $UAV_j$ and leads $\rho_i$ to update its timestamp $T_{SYS}$ with a received timestamp $t_j$. 

\subsubsection*{Session Key Generation}
The shared session key $K_S =  HMAC_{K'_{ij}}( t_j || r_{i} || W_{S_j})  $ is locally generated in both $UAV_j$ and $\rho_i$ using parameters exchanged during the mutual authentication phase in steps $d_{26}$ and $d_{34}$, respectively. It should be noted that, a session key $K_S$ only serves to encrypt data exchanged, if need arise. However, the generated shared key $K_S$ plays no role during the next authentication sessions and is only valid during the respective time frame, hence a session key. The proposed protocol does not require synchronizing or updating parameters between authentication sessions.


\subsection{Serverless Secure Tag Search Protocol}
\label{search-prot}
RFID tag search protocol allows an UAV to securely search for a particular tag among a group of tags within its vicinity, authenticate the tag and initiate a secure data exchange session. RFID tag search functionality is a basic and invaluable tool for efficiently searching among a large amounts of tags~\cite{tan2007severless} without the need to authenticate all tags in the vicinity prior to finding the right one. The tag search protocol minimizes the time to search for a known tag within a group of tags.

\subsubsection*{Authentication}
When $UAV_j$ wants to search for a specific tag with a temporary identity $temp_{ij}$ from the list of tags $L_j$, it calculates $H_{ij} = HMAC_{K_{ij}}(t_j)$ where $t_j$ is the $UAV_j$'s current timestamp and $K_{ij}$ is the key corresponding to a tag with identity $temp_{ij}$. $UAV_j$ broadcasts message $A$ containing $W_{S_j}$, $AR_{ij}, H_{ij}$ and $t_j$ to all tags in the vicinity.

After receiving message~\textit{$A$}, a tag $\rho_i$ validates the parameters received, calculates its temporary key $K_{ij}$ and checks whether it is the intended recipient tag by calculating and comparing $H'_{ij} == H_{ij}$. If it is indeed the intended tag and the values are correct, the tag authenticates the $UAV_j$ and $\rho_i$ updates its timestamp $T_{SYS}$ before replying with a challenge $V_{ij}$ and $r_i$ to $UAV_j$. The other tags do not respond to the query.

Upon receipt of message~\textit{$B$}, $UAV_j$ verifies $V_{ij}$. If $V_{ij}$ is valid, $UAV_j$ authenticates $\rho_i$.

\subsubsection*{Session Key Generation}
$UAV_j$ and tag $\rho_i$ compute a shared key $K_S$ using parameters from both parties in steps $e_{23}$ and $e_{18}$, respectively. $K_S$ is used to securely exchange data between $UAV_j$ and $\rho_i$ using an encryption scheme which is out of the scope of this paper.



\section{Performance Analysis}
\label{sec:Performance Analysis}
\label{prot-analysis}
This section analyses the performance of our proposed authentication and search protocols relative to other protocols.\\
It is worth noting that the \textit{Computational Cost} of both UAVs and tags for both protocols is lightweight and compatible with resource constrained devices since they only need to have capability to execute basic primitives such as concatenation, comparison, and HMAC.

\subsection{Performance Analysis for the Authentication Protocol}
\label{prot-analysis-sub}
Table~\ref{table:auth-comparison-performance} compares our mutual authentication protocol to other similar protocols. Hoque et al.~\cite{hoque2010enhancing} protocol performs authentication using four messages, this translates to energy overhead due to sending and receiving operations. Our protocol is attractive as it supports mutual authentication in three messages. Finally, while Lee et al.~\cite{lee2012server}, Hoque et al.~\cite{hoque2010enhancing} and Abdolmaleky et al.~\cite{abdolmaleky2015strengthened} use hash function in their implementation, our protocol integrates a HMAC function for providing a higher security. However, the HMAC function in our protocol can be easily replaced by any secure hash or PRNG function that suit the tag's requirement and the proposed security and performance properties of our protocol will still hold.

\begin{table}[h]
\caption{Performance comparison between our serverless authentication and similar protocols}
\centering
\begin{footnotesize}

\begin{tabular}{ >{\arraybackslash}m{0.9in} > {\centering\arraybackslash}m{0.4in}  >{\centering\arraybackslash}m{0.4in}  >{\centering\arraybackslash}m{0.4in}  >{\centering\arraybackslash}m{0.5in} }
\toprule
Criteria & Lee et al.~\cite{lee2012server} & Hoque et al.~\cite{hoque2010enhancing} & Abdolmaleky et al.~\cite{abdolmaleky2015strengthened} & Our Protocol \\
\midrule
Computation cost  & 4 hash& 2 hash / 1 PRNG & 4 hash & 3 hmac / 1 PRNG      \\
Total messages     & 3  & 4    & 3  & 3 \\
Storage cost (bits)& 896 & 1024 & 1024 & 864 \\

\bottomrule
\end{tabular}
\end{footnotesize}
\label{table:auth-comparison-performance}
\end{table}

\noindent
\textit{Communication Cost}: In our protocol, the tag $\rho_i$ sends 288 bits (36 bytes) and receives 512 bits (64 bytes) of data during communication. \\\\
\textit{Storage Cost}: At the tag, our protocol uses 160 bits for storing timestamp $T_{SYS}$ and tag's identifier $id_i$ together with an additional 480 bits during operation for storing $r_j, K_{ij}$ and $V_{ij}$ which makes a total of 640 bits or 80 bytes. At the peak moment, just before sending message \textit{B}, the tag must store a total of 864 bits or 108 bytes. The $UAV_j$ storage demands vary depending on the number of tags it is allowed to authenticate at a time, that is the number of tag parameters contained in list $L_j$.\\

\subsection{Performance Analysis for the Secure Tag Search Protocol}
\label{search2-protocol-security-analysis}
The similarities between RFID secure tag search protocol and mutual authentication protocol lead to similarities in performance and security properties. As such, this section discusses only a few properties that differ from those discussed in section~\ref{prot-analysis-sub}. 

The performance comparison between our secure tag search protocol with other similar protocols is given in Table~\ref{table:search-comparison-performance}. 
\begin{table}[h]
\caption{Performance comparison between our tag search protocol and similar protocols}
\centering
\begin{footnotesize}

\begin{tabular}{ >{\arraybackslash}m{0.9in} > {\centering\arraybackslash}m{0.4in}  >{\centering\arraybackslash}m{0.4in}  >{\centering\arraybackslash}m{0.4in}  >{\centering\arraybackslash}m{0.5in} }
\toprule
Criteria & Jeon et al.~\cite{jeon2014ultra} & Hoque et al.~\cite{hoque2010enhancing} & Xie et al.~\cite{xie2014rfid} & Our Protocol \\
\midrule
Computation cost     & 4 PRNG  & 2 hash / 3 PRNG  & 4 hash  & 3 hmac / 1 PRNG \\
Total messages       & 3       & 4                & 3       & 2               \\
Storage cost (bits)  & 896     & 1024             & 1026    & 896            \\
\bottomrule
\end{tabular}
\end{footnotesize}
\label{table:search-comparison-performance}
\end{table}

\noindent \textit{Communication Cost}: Tag search protocol exchanges only two messages, one from each party where $\rho_i$ sends 288 bits (36 bytes) and receives 384 bits (48 bytes) of data.
\\\\
\textit{Storage Cost}: 
The peak storage for the tag search protocol is the moment just before the tag sends message \textit{B}. The total storage space required on the tag is 896 bits or 112 bytes. This corresponds to the total size of $K_{ij}$, $V_{ij}$,  $W_{S_j}$, $AR_{ij}$, $t_j$, $r_i$, $id_i$, $T_{SYS}$, and $K_S$.  

\section{Security and Privacy Analysis}
\label{sec:Security and Privacy Analysis}
\label{prot2-sec-priv-analysis}
This section analyses the security of our proposed mutual authentication and secure tag search protocols using relevant threat and attack models put forth in section~\ref{threat-models}. 

\subsection{Security Analysis for the Authentication Protocol}
\label{prot-analysis-authprot}
As depicted in Table~\ref{table:auth-comparison}, our protocol is the only one that guarantees the privacy of the tags and its secrets when the reader is compromised. The rest of the protocols fail to revoke or change the information granted to the reader after the initial authorization phase. Moreover, in Lee et al.'s~\cite{lee2012server} proposed protocol, the tag always responds with a constant value, which makes it traceable in all communications with the same reader. Likewise, Hoque et al.'s~\cite{hoque2010enhancing} protocol does not provide mutual authentication between the tag and UAV, hence diminishing the level of trust and security of the protocol.\\

\begin{table}[h]
\caption{Security comparison between our serverless authentication protocol and similar protocols}
\centering
\begin{footnotesize}
\begin{tabular}{ >{\arraybackslash}m{1.0in}  >{\centering\arraybackslash}m{0.4in}  >{\centering\arraybackslash}m{0.4in}  >{\centering\arraybackslash}m{0.4in}  >{\centering\arraybackslash}m{0.4in} }
\toprule
Security requirement & Lee et al.~\cite{lee2012server} & Hoque et al.~\cite{hoque2010enhancing} & Abdolmaleky et al.~\cite{abdolmaleky2015strengthened} & Our Protocol \\
\midrule
Tag untraceability & No &  No & Yes & Yes\\
Avoid tag impersonation & No & Yes & Yes  & Yes\\
Avoid replay attack & No & Yes & No & Yes \\
Mutual authentication & Yes & No & Yes & Yes\\
Reader compromise resistance  & No & No & No & Yes\\
\bottomrule
\end{tabular}
\end{footnotesize}
\label{table:auth-comparison}
\end{table}
\subsubsection*{\textit{Game 1 - $\alpha$ masquerades as UAV}}: Referring to $Game~1$ in section~\ref{threat-models}, $\alpha$'s objective is to send legitimate messages \textit{A} and \textit{C}. That is, $\alpha$ can either crack the key $K_{ij}$ or directly generate a valid message \textit{C} based on sniffed messages \textit{A, B} and \textit{C} of earlier legitimate sessions.\\
One way to crack $K_{ij}$ is to extract from message \textit{B} the values of $K_{ij}$ using public values $r_{i}$, $r_{j}$ and $H_{ij}$. This assumes reversibility of HMAC function, which is contrary to the assumption made in section~\ref{assumptions}.\\
Alternatively,  $\alpha$ may combine messages \textit{A, B} and \textit{C} in order to deduce valuable information and use it to crack the key $K_{ij}$. However, messages \textit{B} and \textit{C} behave as random or pseudo-random strings because they evolve independently from each other as their inputs are different. As such, regardless of the number of sniffed messages \textit{A, B} and \textit{C}, it is infeasible to extract any valuable information, and the game cannot succeed.\\
\subsubsection*{\textit{Game 2 - $\alpha$ creates counterfeit tags}}: In our protocol, we do not consider any hardware-based defences against physical attacks. Hence, $\alpha$ may physically compromise a tag $\rho_i$ and access everything in it, including secret information and the information exchanged with $UAV_j$. To create a fake tag $\rho_x$ and fool $UAV_j$, $\alpha$ must know $\rho_x$'s identity $id_x$. As the identity of each tag is secret, different and unique, $\alpha$ cannot guess the identity of tag $\rho_x$ by knowing the identity of $\rho_i$. Thus, compromising a tag $\rho_i$ does not give $\alpha$ the power to derive other tags in $L_j$, hence $\alpha$ cannot win the game.\\
\subsubsection*{\textbf{Game 3 - $\alpha$ tracks $\rho_i$}}: 
Referring $Game~3$ in section~\ref{threat-models}, $\rho_i$ and $UAV_j$ use random values to generate messages \textit{B} and \textit{C}, respectively. During session $k$, $\rho_i$ responds with messages $B_{1ik}$ and $B_{2ik}$, which  appear random to $\alpha$. Any response from $\rho_1$ is semantically indistinguishable from responses of $\rho_2$, and even to the previously sent responses of $\rho_1$. As such, an adversary $\alpha$ is unable to guess with a probability greater than 0.5 which tag $\rho_i$ sent message \textit{B}.

\subsection{Security Analysis for the Secure Tag Search Protocol}
Table~\ref{table:search-comparison} gives a brief comparison between our protocol and other similar protocols. Our protocol protects tags' identities from adversaries. Moreover, Jeon et al.~\cite{jeon2014ultra} suffers from the replay attack while Hoque et al.~\cite{hoque2010enhancing} proposed protocol does not perform mutual authentication, hence reducing trust between the communicating parties. 

Our tag search protocol, like our proposed authentication protocol, is not vulnerable after reader compromise attacks i.e, the values obtained after compromising the reader cannot be used indefinitely. However, the rest of the protocols i.e., those proposed by Jeon et al.~\cite{jeon2014ultra}, Hoque et al.~\cite{hoque2010enhancing}  and Xie et al.~\cite{xie2014rfid} give away crucial information that cannot be revoked once the reader, UAV in our case, is compromised. The adversary may continually use these values to communicate with the respective tags without the possibility of revoking them. 

\begin{table}[h]
\caption{Security comparison between our tag search protocol and similar protocols}
\centering
\begin{footnotesize}
\begin{tabular}{ >{\arraybackslash}m{1.0in}  >{\centering\arraybackslash}m{0.4in}  >{\centering\arraybackslash}m{0.4in}  >{\centering\arraybackslash}m{0.4in}  >{\centering\arraybackslash}m{0.4in} }
\toprule
Security requirement & Jeon et al.~\cite{jeon2014ultra}  & Hoque et al.~\cite{hoque2010enhancing}  & Xie et al.~\cite{xie2014rfid} & Our Protocol \\
\midrule
Tag untraceability & Yes &  Yes & Yes & Yes\\
Avoid tag impersonation & Yes & No & Yes  & Yes\\
Avoid replay attack & No & Yes & Yes & Yes \\
Mutual authentication & Yes & No & Yes & Yes\\
Reader compromise resistance & No & No & No & Yes\\
\bottomrule
\end{tabular}
\end{footnotesize}
\label{table:search-comparison}
\end{table}
\noindent
\subsubsection*{\textit{Game 1 - $\alpha$ masquerades as an UAV}}: Referring to $Game~1$ in section~\ref{threat-models}, $\alpha$'s objective is to send a valid message \textit{A} to $\rho_i$. The first idea would be that $\alpha$ replays a valid message \textit{A}. However, the message is intended for one specific tag with the key $K_{ij}$, and processing of message \textit{A} by the tag leads to the tag updating its timestamp. As a consequence, assuming that the target tag is in the vicinity of the legitimate UAV when transmitting a valid message \textit{A}, replays remain useless as it will be considered by the target tag as out-of-date.

There are two other alternatives for $\alpha$: cracking the key $K_{ij}$ or generating a valid message \textit{A} based on sniffed messages of earlier valid sessions. For cracking $K_{ij}$, one way is to extract the value of $K_{ij}$ from message \textit{A} or~\textit{B} by reversing the HMAC function with known public values $t_j$ or $r_{i}$. However, this contradicts our assumptions of section~\ref{assumptions}.

Alternatively, $\alpha$ can analyse several valid pairs of messages \textit{A} and \textit{B} to generate a new valid message \textit{A}. However, messages \textit{A} and \textit{B} behave as random or pseudo-random strings due to their random inputs. Thus, it is not possible to guess a new valid message \textit{A}, and the game cannot succeed.\\
\subsubsection*{\textit{Game 2 - $\alpha$ creates counterfeit tags}}: This game is similar to the one analysed in the previous authentication protocol in section~\ref{prot-analysis-authprot}.\\

\subsubsection*{\textbf{Game 3 - $\alpha$ tracks $\rho_i$}}: As our search tag protocol facilitates a legitimate $UAV_j$ to search and communicate to a chosen tag within a  group, it is also an ideal opportunity for $\alpha$ to track a tag and launch attacks. 

However, launching a successful attack means $\alpha$ must link message \textit{B} to a particular tag. As messages \textit{B} coming from $\rho_1$ and $\rho_2$ are semantically indistinguishable due to the random inputs $r_i$ and $r_j$, an adversary $\alpha$ cannot guess with a probability greater than 0.5 which tag $\rho_i$ sent message \textit{B}, and he can not win the game.

\section{Conclusion}
\label{sec:Conclusions}

In this paper we have addressed the problem of inventory and tracking of the RFID tagged assets from different actors (airline companies, passengers, maintenance staff, etc.) in airport scenarios, including planes and the assets they embed. We have put forward an approach that uses a UAV. To achieve this approach, we have proposed two serverless protocols:
\begin{enumerate}[label=\alph*)]
\item an authentication protocol for mass identification of a group of RFID tags within a proximity;
\item a search protocol to interact with a specific RFID tags within the proximity.
\end{enumerate}

We have presented: the inventory and tracking system model; the requirements to ensure performance in terms of computation (we are working with constrained systems); the power-consumption issue; security and privacy goals - assumptions, threat and attack models. We then have described the two serverless protocols that we propose to use and we eventually presented the associated performance, security and privacy analysis. For performance, security and privacy, we have shown that the proposed protocols compare favorably with the relevant literature. Regarding energy consumption, as proposed protocols are designed with lightweight operations, they fulfill the objective.

Following the results presented here, we intend to deploy a prototype of such inventory and tracking system to assess its resilience and real world performance.



\end{document}